\documentclass[aps,prl,twocolumn,showkeys]{revtex4}
\usepackage{amsmath}
\usepackage{amssymb}
\usepackage{graphicx}

\begin{document}
\title{ Exact Fidelity and Full Fidelity Statistics in Regular and Chaotic Surroundings}

\author{Heiner Kohler$^{(1)}$, Hans--J{\"u}rgen Sommers$^{(1)}$, Sven \AA berg$^{(2)}$ and Thomas Guhr$^{(1)}$}

\affiliation{$^{(1)}$Fakult{\"a}t f{\"u}r Physik, Universit{\"a}t
Duisburg-Essen, 47057 Duisburg, Germany. $^{(2)}$ Mathematical Physics,
LTH, Lund University, P.O. Box 118, S-221 00 Lund, Sweden}

\begin{abstract} 
For a prepared state exact expressions for the time dependent
mean fidelity as well as for the mean inverse paricipation ratio are obtained
analytically. The distribution function of fidelity in the long time limit and
of inverse participation ratio are studied numerically and analytically.
Surprising features like fidelity revival and enhanced non--ergodicity are observed. 
The role of the coupling coefficients and of complexity of background is studied as well.
\end{abstract}



\maketitle

In quantum information it is crucial to know how well a prepared
state can be isolated, and how the unavoidable mixing with the
surrounding behaves. The fidelity (often called quantum Loschmidt echo or
survival probability)
describes how the purity of a prepared state decreases due to the
interaction with surrounding states \cite{per84}. It serves as a
benchmark test of prepared states, e.g. a set of qubits in quantum
information \cite{nie00}. Of interest is to describe the
explicit time-dependence of the fidelity, and how the decay depends
on the coupling to the environment, as well as the role played by
the complexity of background states. Time-scales, energy scales
and explicit shapes of these functions are thus frequently studied \cite{gor06,jaq08}. 
Survival probability of a state weakly coupled to a background has 
been subject of considerable research in mesoscopics and in semiclassics
\cite{pri94,gru97,van03,gut09}.

 F(ermis) G(olden) R(ule) predicts an exponential
decay of the prepared state with decay rate $\Gamma = 2\pi \lambda^2 D$, where
$D$ is the mean level spacing of the background and $\lambda$ is the average
coupling strength. Deviations from this behavior become important when
$\Gamma\simeq D$ \cite{pri94,gru97}. Corrections to the FGR, which are similar
to  weak localisation corrections in Quantum Transport can lead to
non--ergodicity, i.~e.~ the prepared state will never decay
completely\cite{pri94,gru97}.
 
The observed sa\-tura\-tion of fide\-lity in the long time limit allows us to connect
it to the I(nverse) P(articipation) R(atio), which is a time independent
quantity. Usually average behavior is considered. But the
fidelity in an explicit situation can deviate much from the average
behavior\cite{gor04a}. When constructing quantum information and other devices, requiring
high fidelity, one is interested in a high probability to have states with fidelity
superior to some minimal value, above which error correction is possible \cite{pre98}. 
In this Letter we therefore study fluctuations of the
fidelity and the full fidelity distribution. We find that the latter in the long
time limit tends to a 
stable distribution. This distribution is different from the (time--independent)
IPR distribution, both coincide 
in the weak coupling limit.
In this limit we provide an analytical solution. This allows us to study how 
the distribution depends on coupling
strength, coupling type as well as on dynamics of the background. 

In addition we provide exact solutions for the
mean fidelity for all times and all coupling strength as well as for the mean
IPR. We report on nonperturbative features such as a fidelity recovery and
enhanced non--ergodicity.   

We model the coupling between the prepared state
with surrounding complexity by the Hamiltonian
\begin{eqnarray}
H_{\lambda} &=&  H_s+H_b + \lambda V  \ = \ E_s | s \rangle \langle s
|\nonumber\\
            &&
+ \sum_{\nu=1}^N E_\nu | b_\nu \rangle \langle b_\nu | + \lambda
  \sum_{\nu=1}^N \left( V_\nu | s \rangle \langle b_\nu | + h.c.
\right).
\end{eqnarray}
where $H_s$ represents a special, pure state, 
that is coupled to  a background of complex states described by $H_b$, and where
the
coupling, $\lambda V$, is controlled by the sortless parameter
$\lambda$. Without loss of generality we may put the unperturbed energy of the
special state to zero, $E_s=0$. 

The Schr{\"o}dinger equations for the uncoupled Hamiltonians are
\begin{equation}
\label{SE1}
H_s|s \rangle = 0 \mbox{\hspace{7mm}and\hspace{7mm}}
H_b|b_\nu\rangle = E_\nu |b_\nu\rangle.
\end{equation}
The eigenvalue problem for the coupled Hamiltonian is
\begin{equation}
\label{SE2}
H_{\lambda}|n\rangle = E_n|n\rangle \ .
\end{equation}
The $N+1$ eigenfunctions are expressed in a basis of the special
state and the background states
\begin{equation}
| n \rangle = c_{ns} | s \rangle + \sum_{\nu=1}^{N} c_{n \nu}
|b_{\nu} \rangle,
\end{equation}
where
$c_{ns}=\langle n | s \rangle$ and $c_{n\nu}=\langle n | b_{\nu} \rangle$.
We model the complex surrounding of background states by random
matrix theory, and describe generic chaotic states with an ensemble
of Gaussian random matrices that can either show time-reversal invariance
(GOE, $\beta=1$) or not (GUE, $\beta=2$). Regularity of the background states is
modeled
by assuming Poisson statistics for $H_b$. In all cases the spectrum
is unfolded so the mean level spacing equals one, $D=1$, at
least in a surrounding of the special state. This implies that the
energy scale, including the coupling strength, is always expressed
in units of the mean level spacing, while the time scale
is expressed in units of the Heisenberg time,
$\tau_H=\hbar/D$. In the following $\hbar=1$.

Matrix elements of the operator $V$ between the special state and
the complex surrounding, $V_\nu=\langle s | V | b_\nu \rangle$, are
taken as Gaussian distributed random numbers with zero mean and
variance one. As we will see, it is important to distinguish 
between real coupling ($V_\nu\in {\mathbb R}$) and complex coupling ($V_\nu\in
{\mathbb C}$). 
The size $N$ of the Hilbert space describing the
complex surrounding is in principle infinite in RMT. In the
numerical simulations it is taken sufficiently large to achieve
convergence.

We first focus on fidelity decay. The fidelity amplitude 
$f_\lambda (t) = \langle \Psi_0 (t) | \Psi_\lambda (t) \rangle$ is the 
overlap, between the pure state, time developed under the influence of the
pure Hamiltonian, $ | \Psi_0(t) \rangle =
\exp\left(-i(H_s+H_b) t \right) | \Psi_0(0) \rangle $,
and under the influence of the perturbed Hamiltonian, $|
\Psi_\lambda(t) \rangle = \exp\left( -i H_{\lambda}t
\right) | \Psi_0(0) \rangle$. It is a measure of how the pure initial state,
$|\Psi(0)\rangle$, gets disturbed or mixed due to the
(unavoidable) coupling to the complex surrounding a time $t$ later.
We are interested in the decay of the special state $|s\rangle$,
i.~e.~$|\Psi(0)\rangle =  |s\rangle$. We expand it in a basis of eigenstates to
$H_\lambda$. The fidelity amplitude becomes
\begin{equation}
f_{\lambda} (t) =  \sum_{n} | c_{ns} |^2 \cdot \exp \left(iE_n t \right) .
\end{equation}
It is seen that $f_\lambda (t)$ is the Fourier transform of the local
density of states (LDOS),
\begin{equation}
\rho(E)=\sum_{n} |c_{ns}|^2 \delta (E-E_n).
\end{equation}
The smooth part of the LDOS follows a  Breit-Wigner
distribution, with the width given by
$\Gamma=2\pi\lambda^2$, as obtained from FGR under very general assumptions. 
Therefore the mean fidelity amplitude of the special initial 
state $|s\rangle$ will unavoidably decay exponentially
\begin{equation}
\label{Fidamp}
\overline{f_{\lambda} (t)} = \exp (-\Gamma t/2) \ ,
\end{equation}
where the bar denotes average over background and over coupling matrix elements.

The fidelity (survival probability), of the special state is defined as,
$F_\lambda(t)=|f_\lambda(t)|^2$. In a Drude--type approximation for the mean
fidelity  
\begin{equation}
\label{assumption}
\overline{F_{\lambda}}\ =\ \overline{f_{\lambda}}^2 = \exp(-\Gamma t)\  
\end{equation}
FGR is recovered. This result is also obtained by second order perturbation
theory.
We write fidelity
as the sum  
\begin{equation}
F_{\lambda} (t) =  {\rm IPR} + F_{\rm fluc}(t) \ , \label{fluc}
\end{equation} 
of a constant term and a term which is fluctuating on a timescale comparable to
Heisenberg time. The constant term 
\begin{equation}
{\rm IPR} =  \sum_{n} | c_{ns} |^4 = D \int dE \rho^2(E) ,
\end{equation} 
is the inverse participation ratio of the special state in the basis of the
eigenvectors of the full Hamiltonian. The fluctuating term \begin{equation}
F_{\rm fluc}(t) \ =\  2\sum_{n,m} | c_{ns} |^2 | c_{ms} |^2 \cos((E_n-E_m) t)
\end{equation} 
vanishes, if we average fidelity over a time window, large compared with
Heisenberg time.  Using a Breit--Wigner distribution for $\overline\rho(E)$ and
the Drude approximation Eq.~(\ref{assumption}), $\overline{{\rm IPR}_\lambda} = D/(\pi\Gamma)$
is found \cite{gru97}, which can obviously not hold for small $\Gamma$.  
 
For complex coupling we were able to calculate $\overline{F_{\lambda} (t)}$
exactly for a regular and for a GOE/GUE background. For a regular
background the result \cite{koh09b} is given by
 \begin{eqnarray}
 \label{F2}
\overline{F_\lambda(t)}&=1+ & \frac{\lambda}{2\sqrt{\pi}}\int\limits_{0}^{1}
\frac{dx}{\sqrt{x}} 
                               e^{-\frac{x\pi^2\lambda^2}{4(1-x)}} \nonumber\\
                           && 
\left\{\frac{\pi}{\sqrt{1-x}}\left(e^{-\frac{t^2\lambda^2}{x}}
                              \cosh\left(\frac{\pi\lambda^2 t
}{\sqrt{1-x}}\right)-1\right)\right.\nonumber\\ 
                           &&  \left.-\frac{2 t}{x}e^{-\frac{t^2\lambda^2}{x}}
                               \sinh\left(\frac{\pi\lambda^2 t
}{\sqrt{1-x}}\right)\right\}\ .
\end{eqnarray} 
For a GUE/GOE background the corresponding more complicated expressions can be found in
\cite{koh09b}.
For small times this function decays exponentially according to the FGR law.
Surprisingly, after some characteristic time fidelity reaches a minimum and
increases afterwards to a $\lambda$ dependent saturation value
$\overline{F_\lambda(\infty)} =\overline{{\rm IPR}_\lambda} $. We find for a regular
background
 \begin{eqnarray}
 \label{F3}
\overline{{\rm IPR}_\lambda}&=& 1-\frac{\sqrt{\pi}^3\lambda}{2}{\rm
D}\left(\frac{\pi\lambda}{2}\right) \  ,
\end{eqnarray}
with ${\rm D}(\omega) = \exp(\omega^2){\rm erfc}(\omega)$. $\overline{{\rm
IPR}_\lambda}$ is a monotonously decreasing function. For small coupling the
saturation value behaves as $\overline{{\rm IPR}_\lambda}$ $\simeq$ $
1-\pi^{3/2}\lambda$. For large values of $\lambda$ it decays algebraically as
$\simeq 2/(\pi^2\lambda^2)$ which is four times the value, obtained by the Drude
approximation, Eq.~(\ref{assumption}). This is a striking enhancement of
non--ergodicity. The corresponding expresssion for a GUE background is 
  \begin{eqnarray}
  \label{F4}
 \overline{F_\lambda(\infty)}&=&1-\pi^2\lambda^2-\frac{\sqrt{\pi}^3\lambda}{2}
\left(1-2\pi^2\lambda^2\right)
                                 {\rm D}\left(\pi\lambda\right).
 \end{eqnarray}
This function decays algebraically as $\simeq 1/(\pi^2\lambda^2)$ for large
$\lambda$ which is twice the value, predicted by the Drude approximation. 

\begin{figure}
\includegraphics[width=7cm,clip=true]{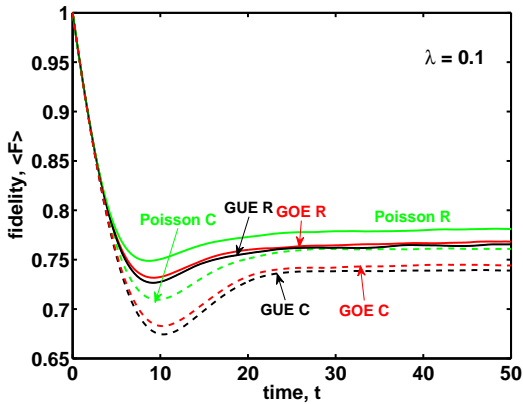}
 \caption{(Color online): Evolution in time of fidelity for coupling strength $\lambda=0.1$. The
full lines describe top down fidelity decay for 
real coupling to a Poissonian (green), to a GOE (red) and to a GUE
(black) background. The dashed lines describe top down fidelity decay for
complex coupling to a Poissonian (green), to a GOE (red) and to a
GUE (blue) background.}
\label{fig1}
\end{figure}

In Fig.\ref{fig1} fidelity $\overline{F_\lambda(t)}$ is plotted for
$\lambda=0.1$ for different complexity of the background. For complex coupling
the curves are obtained from Eq.(\ref{F2}) respectively from the corresponding
expressions taken from Ref.~\cite{koh09b}. For real coupling the curves are
obtained by Monte Carlo simulations. It is seen that in all three cases fidelity
reaches a minimum and saturates afterwards at a finite value. 
Fidelity decay is stronger for a chaotic background than for
a regular background. This is in accordance with the original perturbative
arguments by Peres \cite{per84}.  

Quite remarkably the decay of fidelity is much less sensitive to the complexity
of the background than to the structure of the coupling. There is practically no
difference between a time reversal invariant chaotic background and a background
with broken time--reversal invariance. Nevertheless the difference between a
real coupling of the special state to the background and a coupling which breaks
time reversal symmetry is sizeable. For one reason, because the return probability 
from the background into the special state is suppressed by a
coupling, which breaks time reversal invariance.

\begin{figure}
\includegraphics[width=7cm,clip=true]{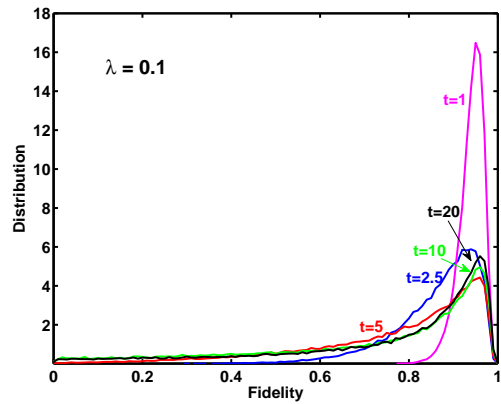}
 \caption{(Color online): Fidelity distributions at times $t=1$ (pink), $t=2.5$ (blue), $t=5$ (red), $t=10$ (green) 
and  $t=20$ (black) for coupling strength $\lambda=0.1$. For times $t>20$
the distribution is stable.}
\label{fig2}
\end{figure}

\begin{figure}
\includegraphics[width=7cm,clip=true]{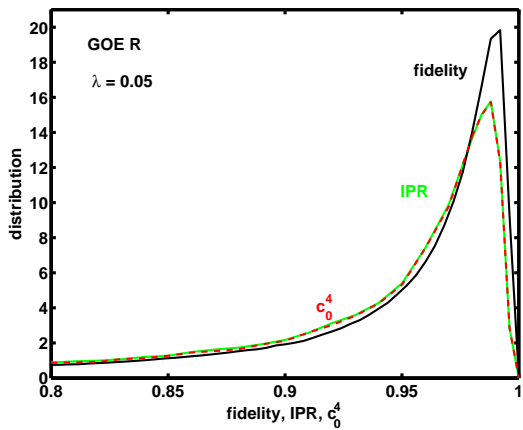}
 \caption{(Color online): Comparison of the distributions $P_0$ obtained from Eq.~(\ref{DisGOE}) (dashed red line) with $P_{\rm
IPR}$ (full green line) and $P_{\rm F}(c,\infty)$ (full black line) obtained from Monte--Carlo simulations for
a coupling constant $\lambda=0.05$.}
\label{fig3}
\end{figure}

We now turn to full fidelity statistics. We introduce the full distribution
function
\begin{equation}
P_{\rm F}(c,t) \ =\ \overline{\delta(c-F_{\lambda}(t))} \ .
\end{equation}
The distribution of the fidelity $P_{\rm F}(c,t)$  is calculated numerically and
plotted for different times in Fig.~\ref{fig2}. A saturation of the distribution
function is
found for times larger than a ($\lambda$--dependent) saturation time. In
Fig.~\ref{fig2} this saturation time is about $20$ times Heisenberg time. 
The saturated distribution is shown in Fig.~\ref{fig3} for GOE statistics and 
for coupling strength $\lambda=0.05$. It 
can be compared to the IPR--distribution 
\begin{equation}
P_{\rm IPR}(c) \ =\ \overline{\delta\left(c-\sum_n|c_{ns}|^4\right)} \ .
\end{equation}
We see that the two distributions are similar but different. The reason lies in the
fluctuating term $F_{\rm fluc}$ (see Eq.~(\ref{fluc})).
Although the ensemble average of $\overline{F_{\rm fluc}}$ vanishes, its
variance does not. Ultimately $F_{\rm fluc}(t)$ contributes substantially to the
full fidelity distribution. As a result the stable fidelity distribution and IPR
distribution are different. 

We are interested mainly in cases of high
fidelity, that is, when the coupling strength, $\lambda$, is small. In this limit $P_{\rm F}(c,\infty)$ should be better and better 
approximated  by $P_{\rm IPR}(c)$. In this limit an analytic result 
for $P_{\rm IPR}(c)$ can be obtained.
By solving the Schr{\"o}dinger equations (\ref{SE1}) and (\ref{SE2}) an
expression for the component of the special
state in the eigenstates $|n\rangle$ of $H_\lambda$ is obtained as
\begin{equation}
|c_{ns}|^2 = \left( 1 + \lambda^2
\sum_{\mu=1}^N\frac{|V_{\nu}|^2}{(E_n-E_{\mu})^2} \right) ^{-1}.
\end{equation}
This exact expression for the components contains eigenvalue
solutions, $E_n$, as well as input matrix elements $V_\nu$ and
energies $E_\nu$. For small $\lambda$ we may approximate the exact eigenvalues
$E_n$ by their unperturbed value $E_{\nu(n)}$. If we denote by $|0\rangle$ the
eigenstate of $H_\lambda$ which has evolved from the special state $|s\rangle$,
we see that in this approximation all $|c_{ns}|$ but $|c_{0s}|$ vanish. This
means that for small couplings the IPR will be dominated by only one term
$|c_{0s}|^4$. We define the distribution
\begin{equation}
P_0(c) \ =\ \overline{\delta\left(c- |c_{0s}|^4\right)}\ .
\end{equation}
Then for small $\lambda$, $P_{\rm F}(c,\infty)$ $\simeq $ $P_{\rm IPR}$ and likewise $P_{\rm IPR}$ $\simeq P_0$. 
We could perform the ensemble average of $P_0$ exactly. For detail of the
calculation see \cite{koh09a}.
For a regular surrounding (Poisson statistics) we find:
\begin{equation}
\label{DisPoi}
P_{0}(c)=\frac{1}{4\sqrt[4]{c^3}}
\frac{2 \lambda a_\beta}{(1-\sqrt{c})^{3/2}}\ {\rm e}^{-(\lambda
a_\beta)^2\frac{\pi \sqrt{c}}{1-\sqrt{c}}}.
\end{equation}
where $a_1$ $=\sqrt{2/\pi}$ for real coupling and  $a_2$ $=\sqrt{\pi/4}$ for
complex coupling.
For chaotic surrounding we find for time-reversal symmetry (GOE):
\begin{equation}
\label{DisGOE}
P_{0}(c)= \frac{1}{4\sqrt[4]{c^3}}\sqrt{\frac{\pi^3\lambda^6
c}{2(1-\sqrt{c})^5}}\ {\rm e}^{-X_1}
\left( K_0(X_1)+K_1(X_1) \right)
\end{equation}
where $K_n$ are modified Bessel functions of second kind. For a
chaotic background without time reversal symmetry (GUE) we get:
\begin{equation}
\label{DisGUE}
P_{0}(c)\ =\ \frac{1}{4\sqrt[4]{c^3}}\sqrt{\frac{2\pi\lambda^2}{(1-\sqrt{c})^3}}\ 
{\rm e}^{-X_2}(1+2X_2),
\end{equation}
with
$X_\beta$ $=$ $\frac{\beta^2\pi^2\lambda^2\sqrt{c}}{4(1-\sqrt{c})}$ .
In Fig.~\ref{fig3} we compare the analytical expressions
for the distribution $P_0$ to numerical simulations of $P_{\rm
IPR}$ and of  $P_{\rm F}(c,\infty)$ for 
$\lambda=0.05$ in the case of a GOE background. 
We observe good agreement of the analytically calculated $P_0(c)$ with 
the distribution $P_{\rm IPR}(c)$ obtained by simulations. Both curves are indistinguishable at least in the range $c\geq 0.6$. 
There is a small but notable difference to the distribution $P_{\rm F}(c,\infty)$, which vanishes if we go to smaller values of $\lambda$.   

In conclusion, we studied fidelity decay for a special state coupled to a
regular or chaotic environment. We found a saturation in the long
time limit and a revival.   
In \cite{pro04} a fidelity freeze was predicted. For a purely off--diagonal
perturbation, after an initial decay fidelity freezes on a plateau for some time
and decays afterwards to zero.  In the model considered here, the
perturbation $V$ is purely off--diagonal as well. The saturation, we found here
might thus be considered an extreme case of fidelity freeze. 
The fidelity revival found here is genuinely different to the one reported
earlier \cite{sto04}. There, a satisfactory explanation was given by the spectral rigidity of the GUE/GOE\cite{sto04,koh08}. 
The fact that the revival occurs for a regular background as well encumbers such an explanation in the present case.     

The saturation of fidelity is a direct consequence of the fact that the fidelity
distribution relaxes in the long time limit into a stable distribution. In the small coupling limit it becomes 
the distribution of the IPR. In this limit we found an analytic expression.
Both distributions have a rich structure and are highly sensitive to even small
changes in the coupling strength. Their relation to maximum strength distribution,
introduced recently \cite{bog06,abe08} will be discussed elsewhere
\cite{koh09a}. Their calculation for arbitrary coupling strength is a challenge for the future.  

Our results yield an important benchmark for the decay of a prepared quantum
state. It might be probed numerically and experimentally in chaotic quantum
systems \cite{fei06,baeck08a,baeck08b} or on quantum information devices. One
instance for a possible numerical experiment is the decay rate of a regular
state in  a mushroom billiard due to dynamical tunnelling into the chaotic part
of the phase space.  

\begin{acknowledgments}
We thank B. Gutkin, R. Oberhage, P. Mello and T. H. Seligman for useful
discussions. We acknowledge support from Deutsche Forschungsgemeinschaft by the
grants KO3538/1-2 (HK), Sonderforschungsbereich Transregio 12 (TG, HK, HJS). 
\end{acknowledgments}

\end{document}